\documentclass[11pt]{llncs}

\usepackage{xspace}
\usepackage[english]{babel}
\usepackage[latin1]{inputenc}
\usepackage{float}
\usepackage{subfigure}

\usepackage[svgnames, table]{xcolor}

\usepackage{color, colortbl}

\usepackage{a4wide}

\usepackage{amsmath,amssymb}

\usepackage{graphicx}
\usepackage[export]{adjustbox}
\usepackage{comment}
\usepackage{textcomp}
\usepackage{booktabs}

\usepackage[hyphens]{url}
\usepackage[breaklinks=true,pdftex]{hyperref}  

\def\ie{{i.e.},~}
\def\eg{{e.g.},~}

\def\z3{{\sc Z3}\xspace}

\colorlet{vert}{green!70!black}
\colorlet{rouge}{red!70!black}
\colorlet{orange}{orange!100!black}
\colorlet{bleu}{cyan!80!white!80!black}
\colorlet{gris}{black!10!white}

\usepackage{tikz}
\usetikzlibrary{arrows,backgrounds,calc,trees}

\pgfdeclarelayer{background}
\pgfsetlayers{background,main}

\tikzset{
  inode/.style = {align=center, inner sep=0pt, text centered,
    font=\sffamily, circle, draw=white, black, fill=white, 
    text width=2.1em, very  thick}
  }

\newcounter{mynote}
\newlength\mynotewidth
\newcommand{\FC}[1]{%
  \stepcounter{mynote}%
  \raisebox{.36em}{\colorbox{green!20!white}{\textcolor{black}{\tiny\textsf{\S$^{\themynote}$}}}}%
  \marginpar{%
    \sffamily%
    \hspace*{-0.2cm}
    \begin{tabular}{l}
     \colorbox{green!30!white}{\color{green!80!black}\parbox{\mynotewidth}{\small\bfseries{\S~{\themynote}}\hfill
       [Franck]}} \\
    \colorbox{green!50!white}{\color{black}\parbox{\mynotewidth}{\fontsize{7}{8}\selectfont
        #1}}
    \end{tabular}
  }
}

\newcommand{\JF}[1]{%
  \stepcounter{mynote}%
  \raisebox{.36em}{\colorbox{orange!20!white}{\textcolor{black}{\tiny\textsf{\S$^{\themynote}$}}}}%
  \marginpar{%
    \sffamily%
    \hspace*{-0.2cm}
    \begin{tabular}{l}
     \colorbox{orange!30!white}{\color{orange!80!black}\parbox{\mynotewidth}{\small\bfseries{\S~{\themynote}}\hfill
       [Joanne]}} \\
    \colorbox{orange!50!white}{\color{black}\parbox{\mynotewidth}{\fontsize{7}{8}\selectfont
        #1}}
    \end{tabular}
  }
}

\usepackage{listings}

\lstdefinelanguage{dafny}{
  sensitive=true,
  keywords={},
  otherkeywords={
  <,>, <=, >=, |, ==, :=, int, seq, tail, init, last, first, take
  },
  basicstyle=\fontsize{10}{12}\selectfont\sffamily,
  keywords = [2]{var, function, lemma, method, ghost, if, then, else, ensures, requires, decreases, while, do, return, od, assert, invariant},
  keywordstyle={\bfseries\color{orange}},
  keywordstyle=[2]{\bfseries\color{blue!80!black}},
  identifierstyle=\color{black},
  comment=[l]{//},
  moredelim = [s][\color{gray}\ttfamily]{/**}{*/},
  commentstyle=\color{gray}\ttfamily,
  stringstyle=\color{red}\ttfamily,
  morestring=[b]",
  frame=lines
}

\lstdefinelanguage{python3}{
  language=Python,
  sensitive=true,
  morekeywords={assert},
  numbers=left,
  basicstyle=\fontsize{10}{12}\selectfont\sffamily,
  keywordstyle={\bfseries\color{orange}},
  identifierstyle=\color{blue},
  commentstyle=\color{gray}\ttfamily,
  stringstyle=\color{red}\ttfamily,
  frame=lines
}

\makeatletter
\renewcommand{\paragraph}{%
  \@startsection{paragraph}{4}%
  {\z@}{2.25ex \@plus 1ex \@minus .2ex}{-1em}%
  {\normalfont\normalsize\bfseries}%
}
\makeatother

\AtBeginDocument{}

\title{\LARGE \bf Formal Verification of the Ethereum 2.0 \\ Beacon Chain\thanks{This work was partially supported by the Ethereum Foundation, grant FY20-285, Q4-2020}}

\author{Franck Cassez\inst{1} \and Joanne Fuller\inst{1} \and Aditya Asgaonkar\inst{2}
  \institute{
    ConsenSys\\
    \and 
    Ethereum Foundation\\
    \email{franck.cassez@consensys.net} \hskip 1em \email{joanne.fuller@consensys.net}
    \hskip 1em \email{aditya.asgaonkar@ethereum.org}
  }
}

\makeatletter
\renewcommand{\paragraph}{%
  \@startsection{paragraph}{4}%
  {\z@}{0.9ex \@plus 1ex \@minus .2ex}{-1em}%
  {\normalfont\normalsize\bfseries}%
}
\makeatother

\begin{document}

\pagestyle{plain}
\maketitle

\thispagestyle{empty}

\begin{abstract}
  We report our experience in the formal verification of the reference implementation of the Beacon Chain.
  The Beacon Chain is the backbone component of the new Proof-of-Stake Ethereum 2.0 network: it is in charge of tracking 
  information about the \emph{validators}, 
  their \emph{stakes}, their \emph{attestations} (votes) and if some validators are found to be dishonest, to \emph{slash} them (they lose some of their stakes).   
  The Beacon Chain is mission-critical and any bug in it could compromise the whole  network.
  %
  The \emph{Beacon Chain reference implementation} developed by the Ethereum Foundation is written in Python, and provides a detailed operational description of the state machine each Beacon Chain's network participant (node) must implement. 
  We have formally specified and verified the absence of runtime errors in (a large and critical part of) the Beacon Chain reference implementation using the verification-friendly language Dafny.
  During the course of this work, we have uncovered several issues, proposed \emph{verified} fixes.
  We have also synthesised \emph{functional correctness specifications} that enable us to provide guarantees beyond
  runtime errors. Our software artefact is available at \url{https://github.com/ConsenSys/eth2.0-dafny}.

\end{abstract}

\section{Introduction}\label{sec-intro}

The Ethereum network is gradually transitioning to a more secure, scalable and energy efficient \emph{Proof-of-Stake (PoS) consensus protocol}, known as Ethereum 2.0 and based off GasperFFG~\cite{gasperFFG}.
The Proof-of-Stake discipline ensures that participants who propose (and vote) for blocks are chosen 
with a frequency that is proportional to their stakes.
Another major feature of Ethereum 2.0 is \emph{sharding} which enables to split the main blockchain into
a number of independent and hopefully smaller and faster chains.
The transition from the current Ethereum 1 to the final version of Ethereum 2.0 (Serenity) is planned over a number of years and will be rolled out in a number of phases.  
The first phase, Phase 0, is known as the Beacon Chain. 
It is the backbone component of Ethereum 2.0 as it coordinates the whole network of \emph{stakers} and shards.

\paragraph{\it \bfseries The Beacon Chain.}

The Beacon Chain (and its underlying protocol) is in charge of enforcing consensus, among the nodes participating in the network, 
on the state of the  system.
The participants are called \emph{validators} and their main role is to \emph{propose} and \emph{vote} for new (Beacon) blocks  
to be appended to the blockchain.
The set of validators is dynamic: new validators can register by staking some ETH (Ethereum crypto-currency).
Once registered, validators are eligible to participate and \emph{propose} and \emph{vote} for new blocks (of transactions) to be appended to the blockchain.
The Beacon Chain shipped on December 1, 2020.
At the time of writing (October 14, 2021), close to $250,000$ validators have staked $7,780,000$ ETH (\$30 Billion USD).
Considering the coordination role and the amount of assets managed by the Beacon Chain, it is a mission-critical component of the Ethereum 2.0 ecosystem, and any bug in it could badly impact the network.
At the same time, the operational description of the Beacon Chain in the \emph{reference implementation} is rather complex and provided in a Python-like language.
It is the reference for any Beacon Chain client implementer. As a result, inaccuracies, ambiguities, or bugs in the reference implementation will lead to erroneous and/or buggy clients that can compromise the integrity, or the performance of the network.

\paragraph{\it \bfseries Our Contribution.}
The \emph{Beacon Chain reference implementation}, developed by the Ethe\-reum Foundation, is written in Python, and provides a detailed operational description of the state machine each Beacon Chain's network participant (node) must implement. 
We have formally specified and verified the absence of runtime errors in (a large and critical part of) the Beacon Chain reference implementation using the verification-friendly language Dafny.
During the course of this work, we have uncovered several issues, proposed \emph{verified} fixes, some of which have been integrated in the reference implementation.
We have also synthesised \emph{functional correctness specifications} that enable us to provide guarantees beyond
runtime errors. Our software artefact with the code and proofs in Dafny is freely available at \url{https://github.com/ConsenSys/eth2.0-dafny}.

\paragraph{\it \bfseries Related Work.}
The Ethereum Foundation has supported several proj\-ects related to applying formal methods for the analysis of the Beacon Chain (and other components).
A foundational project\footnote{\url{https://github.com/runtimeverification/beacon-chain-spec}} was undertaken in 2019 by Runtime Verification Inc. and provided a formal  and \emph{executable} semantics in the K framework, to the reference implementation~\cite{k-exec-specs}. 
The semantics was validated and the reference implementation could be tested which resulted in a first set of recommendations and fixes to the reference implementation.
Runtime Verification Inc. have also formally specified and verified (in Coq~\cite{gasper-coq}) the underlying GasperFFG~\cite{gasperFFG} protocol.
Our work complements these formal verification projects.     
Indeed, our objective is to provide guarantees for the \emph{absence of bugs} (runtime errors), and \emph{loop termination} which goes beyond testing.  
We have chosen to use a verification-friendly programming language, Dafny~\cite{dafny-ieee-2017}, as it enables us to write the code in a more developer-friendly manner (compared to K).
However, we have used the code bases from the previous projects to guide us (\eg semantics) during the course of this project.

\section{The Beacon Chain Reference Implementation}\label{sec-ref-implem}
In this section we introduce the system we want to formally verify, what are the potential 
benefits and impacts of such of study, 
and we set out the goals of our experiment. 

\subsection{Scope of the Study}


As a robust decentralised system, the Beacon Chain aims to implement a \emph{replicated state machine}~\cite{Lamport78} that is fault-tolerant to a fraction of unreliable (\eg they can crash) participants.
The replicated state machine is implemented with a number of networked identical state machines running concurrently.
%
This provides redundancy and a more reliable system.
The state of each machine changes on an occurrence of an \emph{event}. 
As the machines operate asynchronously, two different machines may receive different events 
that cannot be totally ordered time-wise. 
This is why before processing an event and changing their states, the state machines run a \emph{consensus protocol} to decide  
which event they should all process next.
The consensus protocol aims to guarantee (under certain conditions) that an agreement will be reached which 
ensures that events are processed in the same order on each machine. 
\begin{quote}
    \em In this project, we are \textbf{not} interested in the verification of the consensus protocol, but rather in the verification of the state machine that is replicated. 
\end{quote}

\subsection{The Beacon Chain Reference Implementation}
The Beacon Chain (Phase 0) reference implementation~\cite{beacon-phase0} describes the \emph{state machine} that every Beacon node (participant) has to implement.
%
%
%
%
The idea is that anyone is allowed to be a participant in the decentralised Ethereum 2.0 ecosystem when it is fully deployed.
However, as the consensus protocol is Proof-of-Stake there must be a mechanism for participants to register and stake, to slash a participant's stake if they are caught\footnote{In a distributed system with potentially dishonest participants, it is not always possible to detect who is dishonest (byzantine). However, sometimes a participant can sometimes be proved to be dishonest.}  misbehaving, \ie not following the consensus protocol, and to reward them  if they are honest.
The Beacon Chain provides these mechanisms.  
It maintains records about the participants, called \emph{validators},
ensuring fairness (each honest participant should have a voting power, for new blocks, related to its stake), and safety (a dishonest participant may be slashed and lose part of their stakes).


The full Beacon Chain (Phase 0) reference implementation~\cite{beacon-phase0} comprises three main sections:
\textbf{1.} the \emph{Beacon Chain State Transition} for the Beacon state machine is the most complex component;
\textbf{2.} The \emph{Simple SerialiZe (SSZ) library} for how to encode/decode (serialise/deserialise) data that have to be communicated over the network between nodes;
\textbf{3.} the \emph{Merkleise library} for how to build efficient encoding of data structures into Merkle trees, and how to use them to verify Merkle proofs;

\paragraph{\it \bfseries  The State Transition.}
The \emph{Beacon Chain state transition} part is the most critical part and
at the operational level the complexity stems from:
\begin{itemize}
    \item time is logically divided into \emph{epochs}, and each epoch into a fixed number of \emph{slots}; the state is updated at each slot;
    \item at the beginning of each epoch, disjoint subsets of validators are assigned to each slot to participate in the block proposal for the slot and attest (vote) for \emph{links} in the chain;
    \item the state updates that apply at an epoch boundary are more complex than the other updates; 
    \item the actual state of the chain is a \emph{block-tree} \ie a tree of blocks, and the canonical chain is defined as a particular \emph{branch in this tree}. How this branch is determined is defined by the \emph{fork choice rule}. 
    \item the \emph{fork choice} rule relies on properties of nodes in the block-tree, namely, \emph{justification} and \emph{finalisation}. The state update describes how nodes in the block-tree are deemed justified/finalised. 
    The rules for justification and finalisation are introduced in a separate document, the GasperFFG~\cite{gasperFFG} protocol.
    
\end{itemize} 

\paragraph{\it \bfseries The SSZ and Merkleise Libraries.}
These components are self-contai\-ned and independent from the state transition.
We used them as a feasibility study and we had verified them before this project started.
We have provided a complete Dafny reference implementation for them in the \texttt{merkle} and
\texttt{ssz} packages~\cite{eth2-dafny}.

\subsection{Motivation for Formal Verification}\label{sec-motivation}

As mentioned previously, the Beacon Chain shipped on December 1, 2020 and 
up to date, $250,000$ validators have staked $7,780,000$ ETH (\$30 Billion USD).
It is clear that any bug, or logical error, could have disastrous consequences resulting is losses of assets, or downtimes which means losses of rewards for the validators. 
There are regular opportunities (forks) to update the code of Beacon Chain nodes.
So continuously running projects like ours is very valuable as what is important is to find and  
fix bugs before attackers can exploit them. 
The operational description of the Beacon Chain in the reference implementation is provided in a Python-like language.
It was written by several reference implementation writers at the Ethereum Foundation and due to its size it is hard for one person to have a complete picture of it.
It is the reference for any Beacon Chain client implementer. 
%
Moreover the reference implementation uses a defensive mechanism against unexpected errors:
\begin{quote}
    \em [S1]``State transitions that trigger an unhandled exception (e.g. a failed assert or an out-of-range list access) are considered invalid. State transitions that cause a uint64 overflow or underflow are also considered invalid.''
\end{quote} 
However this creates a risk that 
errors unrelated to the logic of the state transition function may introduce spurious exceptions. 
At the time of writing, there are at least 4 different Ethereum 2.0 client softwares that are used by validators. Bugs in the reference implementation may be handled differently in the various clients, and in some cases lead to a split in the network\footnote{A network split can be caused if some clients reject a chain that is being followed by the other clients, which leads to a hard fork-like situation.}. 
The correctness of the consensus mechanism is guaranteed for up to $1/3$
of malicious nodes that are nodes deviating from the reference implementation, be it intentionally or unintentionally (\eg because of a bug in the code). 
Hence, we should try to make sure we reduce (buggy) unintentional malicious nodes.
%


\subsection{Objectives of the Study}\label{sec-objectives}

Our goal is to improve the overall safety, readability and usability of the reference implementation.
%
The primary aspect of our project was to make sure that the code was free of runtime errors (\eg over/underflows, array-out-of-bounds, division-by-zero, \ldots). 
This provides more confidence that when an exception occurs and a state is left unchanged as per [S1], the root cause is a genuine problem related to the state transition having been given an ill-formed block: if \verb+state_transition(state,signed_block)+ triggers an exception, it should imply that there is a problem with the \verb+signed_block+ not that some intermediate computations resulted in runtime errors.
A secondary goal was to try and synthesise \emph{functional specifications} from the reference implementation. This can help developers to design tests, and contributes to the specifications being language-agnostic. For instance, it can help write a client in a functional language which results in a more inclusive ecosystem.

\section{Formal Specification and Verification}\label{sec-formal}

In this section we present the challenges of the project, 
motivate our methodology and conclude with our results' breakdown. 



\subsection{Challenges}\label{sec-challenges}

The main challenges in this formal verification project are in the verification of the code of the 
\verb+state_transition+ component of the Beacon Chain.
The  SSZ and Merkleise libraries are much smaller, simpler, and independent components that
can be dealt with separately. 


The reference implementation for the Beacon Chain~\cite{beacon-phase0} introduces data types and
algorithms that should be \emph{interpreted} as Python~3 code.
As a result it may not be straightforward for those who are not familiar with Python to understand 
the meaning of some parts of the code.
More importantly, the reference implementation is not executable and may contain type mismatches, incompatible function signatures, and bugs that can result in runtime errors like 
under-overflows or array-out-of-bounds. 

A typical function in the reference implementation is written as a sequence of control blocks (including function calls) intertwined with \emph{checks} in the form of \verb+assert+ statements. 
The  \verb+state_transition+ function (Listing~\ref{algo-state}) is the component that computes the update of the Beacon Chain's state.
The state (of type \verb+BeaconState+) records some information including the validators' stakes, the subsets of validators (\emph{committees}) allocated to a given slot,
and the hashes\footnote{The actual blocks are recorded in the \texttt{Store} which is a separate data structure.} of the blocks that have already been added to the chain.  
A state update is triggered when a (signed) \emph{block} is added to Beacon Chain.
The state machine implicitly defined by the reference implementation generates sequences of states of the form:
\[
    s_0 \xrightarrow{\ b_0\ } s_1 \xrightarrow{\ b_1\ } s_2 \ldots \xrightarrow{\ b_{n}\ } s_{n+1} \ldots \tag{StateT}\label{eq-st}
\] 
where:
\begin{itemize}
    \item $s_0$ is given (initial values), 
    \item $b_0$ is the \emph{genesis block} and,
    \item  for each $i \geq 1, s_{i+1} = \verb+state_transition(+s_i, b_i\verb+)+$.
\end{itemize}



\begin{lstlisting}[language=python3,xleftmargin=2em,label=algo-state,caption={The state transition function.}]
def state_transition(
    state: BeaconState, 
    signed_block: SignedBeaconBlock, 
    validate_result: bool=True
) -> None:
block = signed_block.message
# Process slots (including those with no blocks) since block
process_slots(state, block.slot)
# Verify signature
if validate_result:
    assert verify_block_signature(state, signed_block)
# Process block
process_block(state, block)
# Verify state root
if validate_result:
    assert block.state_root == hash_tree_root(state)
\end{lstlisting}
There are several challenges in testing or verifying this kind of code:
\begin{itemize}
    \item the functions calls (lines~8, 13) \emph{mutate} the input variable \verb+state+; those functions also call other functions that mutate the state.
    \item the semantics is not fully captured by the Python 3 interpretation because of the defensive mechanism [S1] (Section~\ref{sec-motivation}, page~\pageref{sec-motivation}). 
    \item a \emph{valid} state transition is the opposite of an \emph{invalid} state transition (characterised by [S1]). Determining when a computation is not going to trigger runtime errors or failed \texttt{assert}s is non-trivial. This is due to the use of mutating functions that can contain \verb+assert+ statements on values that are the results of  intermediate computations. 
    \item overall the code in the reference implementation does not explicitly define what \emph{properties} \verb+signed_block+ should satisfy to guarantee that \verb+state_transition(state,signed_block)+ is a valid transition. The semantics of the code is as follows: if an exception occurs in  \verb+state_transition+ with input \verb+signed_block+, then this block must be invalid (assuming \verb+state+ is always valid).
    If the code contains a bug that triggers a runtime error unrelated to 
    \verb+signed_block+ (\eg an intermediate computation that overflows, or an array-out-of-bounds in a sorting algorithm), \verb+signed_block+ is declared invalid (not added to the chain).
    \item as there is no reference \emph{functional specification} it is not immediate to understand when a block is invalid, and to write (unit) tests.
    \item finally the correctness of parts of the code rely on hidden assumptions, \eg the total amount of ETH is $X$ so no overflow should happen.
\end{itemize} 
The challenges pertaining to the SSZ and Merkleise libraries are more manageable. First the reference implementation is shorter. Second even if there is no functional specification available, it is reasonably easy to synthesise them. 
Due to the previous weaknesses, the reference implementation~\cite{beacon-phase0} has been the subject of several informal explainers~\cite{danny-ryan-bc,ben-annotated-eth2,beacon-phase0}.
A formal, executable (and suitable for testing) semantics has been designed using the K-framework~\cite{k-exec-specs}.



\subsection{Methodologies}

\paragraph{\it \bfseries Resource Constraints.} Resource-wise, the timeframe or our project was approximately 8 months (October 2020 to June 2021), with a team of two formal verification researchers (first two co-authors) and one Beacon Chain expert researcher (third co-author).  


\paragraph{\it \bfseries Verification Technique.}

The reference implementation is \textbf{not} the operational description of a distributed system, but rather a sequential state machine, as per~\eqref{eq-st}, Section~\ref{sec-challenges}.
Thus, techniques and tools that are adequate for the goals we set are 
related to \emph{program formal verification}.

There are several techniques to approach program verification, ranging from fully automated (\eg static analysis/abstract interpretation~\cite{cousot-ai-21}, software model-checking~\cite{sw-mc-2009}) to  interactive theorem proving~\cite{concrete-semantics-2014}.
Most static analysers are unsound (they cannot prove the absence of bugs) which disqualifies them for our project.   
It is anticipated that fully automated verification techniques can be effective to detect runtime errors but may have limited applicability to proving functional correctness. 

On the other side of the spectrum, interactive theorem provers offer a complete arsenal of logics/rules that can certainly be used for this kind of projects.
However they usually require encoding the software to be verified in a high-level mathematical language that is rather different to a language like Python. 
The level of expertise/experience required to properly use these tools is also high. Overall this seemed incompatible with our available resources. 

A middle-ground between fully automated and interactive techniques  is deductive verification available in verification-friendly programming languages like Dafny~\cite{dafny-ieee-2017}, Why3~\cite{why3-filliatre}, Viper~\cite{viper-2017} or Whiley~\cite{whiley-setss-2018}.
Deductive verification lets verification engineers \emph{propose} proofs and check them fully automatically.

We opted for Dafny~\cite{dafny-ieee-2017}, an award-winning verification-friendly language.
Dafny is actively maintained\footnote{\url{https://github.com/dafny-lang/dafny}} and under continuous improvement. It offers imperative/object oriented and functional programming styles.
Moreover, some of us had a previous exposure to Dafny (working on the SSZ/Merkleise libraries early in 2020), and we could be fully operational quickly, and it was compatible with our resources.
We are convinced that similar results could be achieved with Why3, Viper or Whiley
but did not have the resources to launch concurrent experiments.

\paragraph{\it \bfseries Verification Strategy.}

Our strategy to write the Beacon Chain reference implementation in Dafny and detect/fix runtime errors, and prove some functional properties is three-fold:
\begin{enumerate}
    \item \textbf{Identify simplifications.} The reference implementation is complex
    and trying to encode it fully in Dafny may result in inessential details hindering our verification progress. One example is the different data types (classes) for \verb+Attestations+. There are several variations of the type \verb+Attestations+ and functions to convert between them. It turns out that for our verification purposes, using \verb+PendingAttestations+ instead of the fully fledged \verb+Attestations+ was adequate. Another example is the abstraction of \emph{hashing} functions. We assumed an \emph{uninterpreted} collision-free hash function as we did not aim to prove any probabilistic properties involving this function.

    \item \textbf{Translate the reference implementation in Dafny.} 
    This helped the formal verification researchers to familiarise themselves with the reference implementation. During this phase, we focussed on adding \emph{pre} and \emph{post} conditions to the functions of the reference implementation to guarantee the absence of runtime errors. We were also able to prove some 
    interesting invariants: the data structure that contains 
    the block-tree is indeed a \emph{well-formed tree}. This structure is 
    implemented with links from nodes to their parent (where \verb+null+ is a possible parent in the code). The invariant states that the block-tree that is built with the \verb+state_transition+ function satisfies: $i)$ the set of ancestors of any block contain blocks with strictly smaller \emph{slot} number and is finite (no cycles) $ii)$ the set of ancestors of any block in the block-tree always contains the genesis block (with slot 0).  
    
    \item \textbf{Synthesise functional specifications.} In the last phase, we synthetised functional specifications for each function in the reference implementation. We proved that each function in the reference implementation satisfied its functional specification. This enabled us to prove more complex properties as we could do the formal reasoning and proofs on the functional specifications and the results would carry over to the reference implementation. This was an effective solution to be able to prove properties of the reference implementation with lots of \emph{mutations} without having to embed them deep in the proofs.   
\end{enumerate}

\subsection{Results' Breakdown}

The complete code base is freely available in~\cite{eth2-dafny}.
There are several resources apart from the verified code: a Docker container to batch verify the code, some notes and videos to help navigate the
Dafny specifications.

\paragraph{\it \bfseries Coverage.}
We estimated that we have verified 85\% of the reference implementation. The remaining 15\% are simplifications \eg data types, or using a fixed set of validators instead of a dynamic set.
Adding the remaining details to the released version would require a substantial amount of work and at the same time it seems that the likelihood of finding
new issues is low. 
Since the Beacon Chain has shipped in December 1, 2020, only a few minor issues have been uncovered and promptly fixed which seems to confirm the previous claim.

\paragraph{\it \bfseries Absence of Runtime Errors.}
All of the functions we have implemented in Dafny come with annotations in the form of pre (\verb+requires+) and post (\verb+ensures+)  conditions that are verified, including \emph{loop termination}. The Dafny version of function \verb+state_transition+ is given in Listing~\ref{algo-dafny1}.
The functions \verb+process_slots+ and \verb+process_block+ are written in a similar style.
The Dafny verifier enforces the absence of runtime errors like division by zero, under/overflows, array-out-of-bounds. It follows that our code base is provably free of this kind of defect.
Moreover, additional checks can be added like the \verb+assert+ statement at line~33.
We have added all the \verb+assert+ statements from the reference implementation and proved that they could not be violated. This requires adding suitable pre-conditions.

\begin{lstlisting}[language=dafny,numbers=left,xleftmargin=2em,label=algo-dafny1,caption={Dafny version of \texttt{state\_transition}}]
method state_transition(s:BeaconState,b:BeaconBlock) 
                                    returns (s': BeaconState)
  //  A valid state to start from
  requires |s.validators| == |s.balances| 
  requires is_valid_state_epoch_attestations(s)
  //  b must a block compatible with s
  requires isValidBlock(s, b)
  //  Functional correctness
  ensures s' == updateBlock(forwardStateToSlot(nextSlot(s),b.slot),b)
  //  Other post-conditions
  ...
  ensures s'.slot == b.slot
  ensures s'.latest_block_header.parent_root  == 
      hash_tree_root(
          forwardStateToSlot(nextSlot(s), b.slot)
          .latest_block_header
      )
  ensures |s'.validators| == |s'.balances|
  ...
{
    // Finalise slots before b.slot.
    s' := process_slots(s, b.slot);
    
    //  Process block and compute the new state.
    s' := process_block(s', b);  

    //  Verify state root (from eth2.0 specs)
    assert (b.state_root == hash_tree_root(s'));
}  
\end{lstlisting}

\paragraph{\it \bfseries Functional Correctness.}
Beyond the absence of runtime errors, we have also synthesised \emph{functional specifications} based off the reference implementation code.
For instance we have decomposed the state update in \verb+state_transition+ into a sequence of simpler steps, 
\texttt{updateBlock}, \texttt{forwardStateToSlot}, \texttt{nextSlot} and proved that the result is a composition of these functions.
This provides more confidence that the code is functionally correct as our decomposition specifies smaller changes in the state. It also enables us to prove properties on the functional specifications and transfer them to the imperative version of the code.

\paragraph{\it \bfseries Impact of our Project.}
During the course of this projects we have reported several issues, some of them bugs ($3$), some of them need for clarifications ($5$) in the reference implementation. 
The issues we have uncovered are tracked in the \emph{issues tracker} of our github repository.
Some of the bugs we reported have been fixed and our \emph{clarifications} category has led to several improvements in the writing of the reference implementation.
Moreover, we have provided a fully documented version of the reference implementation in Dafny.
The Dafny code contains clear pre and post conditions that can help developers understand the effect of a function and can be used to write unit tests.



\paragraph{\it \bfseries Statistics.}

Table~\ref{tab-stat1}, page~\pageref{tab-stat1}, provides some insights into the actual code, per file.
We have tried to keep the size of each file small and provide optimal modularity in the proofs.
The files in the packages fall into one of the three categories: \texttt{file.dfy} is the reference implementation (Python) translated into Dafny; \texttt{file.s.dfy} contains the functional specifications we have synthesised and 
\texttt{file.p.dfy} any additional proofs (Lemmas) that are used in the correctness proofs.
It is hard to estimate the \emph{lines of code to lines of proofs ratio} for many reasons: $i)$  it is not always possible to locate all the proofs in a separate unit (e.g. a module in Dafny), as this can create circular dependencies. 
\colorlet{proof}{green!70!white}
\colorlet{spec}{yellow}

It follows that counting lines of proofs as lines in the Lemmas is not an accurate measure; $ii)$ in some of the proofs, we have, on purpose, provided redundant hints. 
As a result some proofs can be shortened but this may be at the expense of readability (and verification time).    
For this project, a conservative (and empirical) \emph{lines of code to lines of proofs ratio} seems to be around $1$ to $7$.

\begin{table}[btp]
    \normalsize
    \centering 
\begin{tabular}{llrrrrrr}
\toprule
                       Files &                        Package &  \#LoC &  Lem. &  Imp. &  \#Doc &  $\frac{\#Doc}{\#LoC}$ (\%) &  Proved \\
\midrule
\rowcolor{proof}  ActiveValidatorBounds.p.dfy &                        beacon &    52 &         3 &                0 &             29 &             56 &       3 \\
        BeaconChainTypes.dfy &                        beacon &    54 &         0 &                0 &            171 &            317 &       0 \\
                 Helpers.dfy &                        beacon &  1003 &         9 &               89 &            670 &             67 &      98 \\
\rowcolor{proof}              Helpers.p.dfy &                        beacon &   136 &        13 &                0 &            114 &             84 &      13 \\
\rowcolor{spec}               Helpers.s.dfy &                        beacon &   136 &         9 &                6 &             67 &             49 &      15 \\\hline
       AttestationsTypes.dfy &           beacon/attestations &    30 &         0 &                0 &             68 &            227 &       0 \\ \hline
              ForkChoice.dfy &             beacon/forkchoice &   229 &         3 &               15 &            172 &             75 &      18 \\
         ForkChoiceTypes.dfy &             beacon/forkchoice &     9 &         0 &                0 &             17 &            189 &       0 \\ \hline
                  Crypto.dfy &                beacon/helpers &     7 &         0 &                1 &              3 &             43 &       1 \\ \hline
         EpochProcessing.dfy &        beacon/statetransition &   384 &         0 &               14 &            127 &             33 &      14 \\
         
\rowcolor{spec}        EpochProcessing.s.dfy &        beacon/statetransition &   398 &        24 &                0 &            336 &             84 &      24 \\
       ProcessOperations.dfy &        beacon/statetransition &   361 &         0 &               11 &            119 &             33 &      11 \\
       \rowcolor{proof}  ProcessOperations.p.dfy &        beacon/statetransition &   160 &        10 &                0 &             74 &             46 &      10 \\
     \rowcolor{spec}  ProcessOperations.s.dfy &        beacon/statetransition &   410 &        12 &                6 &            137 &             33 &      18 \\
         StateTransition.dfy &        beacon/statetransition &   215 &         0 &                8 &            126 &             59 &       8 \\
         \rowcolor{spec} StateTransition.s.dfy &        beacon/statetransition &   213 &        11 &                1 &            100 &             47 &      12 \\ \hline
              Validators.dfy &             beacon/validators &    11 &         0 &                0 &             53 &            482 &       0 \\ \hline
               Merkleise.dfy &                        merkle &   504 &         9 &               18 &            135 &             27 &      27 \\ \hline
            BitListSeDes.dfy &                           ssz &   262 &         7 &                3 &             64 &             24 &      10 \\
          BitVectorSeDes.dfy &                           ssz &   155 &         4 &                3 &             53 &             34 &       7 \\
               BoolSeDes.dfy &                           ssz &    22 &         0 &                2 &              3 &             14 &       2 \\
            BytesAndBits.dfy &                           ssz &    90 &         7 &                6 &             44 &             49 &      13 \\
               Constants.dfy &                           ssz &   104 &         0 &                0 &             36 &             35 &       0 \\
                IntSeDes.dfy &                           ssz &   130 &         2 &                2 &             20 &             15 &       4 \\
               Serialise.dfy &                           ssz &   514 &         3 &                5 &             36 &              7 &       8 \\ \hline
                DafTests.dfy &                         utils &    62 &         0 &                4 &             25 &             40 &       4 \\
               Eth2Types.dfy &                         utils &   227 &         1 &                3 &             77 &             34 &       4 \\
                 Helpers.dfy &                         utils &   220 &        11 &                3 &            103 &             47 &      14 \\
             MathHelpers.dfy &                         utils &   293 &        18 &                6 &            105 &             36 &      24 \\
             NativeTypes.dfy &                         utils &    28 &         0 &                0 &             13 &             46 &       0 \\
          NonNativeTypes.dfy &                         utils &     8 &         0 &                0 &              6 &             75 &       0 \\
              SeqHelpers.dfy &                         utils &    69 &         8 &                2 &             58 &             84 &      10 \\
              SetHelpers.dfy &                         utils &    74 &         6 &                0 &             50 &             68 &       6 \\
\rowcolor{gray!50!white}  \textbf{TOTAL} & &  \textbf{6570} &       \textbf{170} &              \textbf{208} &           \textbf{3212} &             
\textbf{49} &     \textbf{378} \\
\bottomrule
\end{tabular}


\caption{Statistics. A \colorbox{spec}{file} providing functional specifications. A \colorbox{proof}{file} providing proofs (lemmas in Dafny). \textbf{\#LoC} (resp. \textbf{\#DoC}) is the number of lines of code (resp. documentation), \textbf{Lem.} the number of proper lemmas, \textbf{Imp.} the number of proved imperative functions with pre/post conditions.}
\label{tab-stat1}
\end{table}



\section{Findings and Lessons Learned}\label{sec-findings}
%

During the course of our formal verification effort we found subtle bugs and also proposed some 
clarifications for the reference implementations.
In addition, our work was the opportunity to start some discussions about how to improve the readability 
of the reference implementation, \eg by using pre and post conditions rather than \verb+assert+ statements.
In this section we provide more insights into some of the main issues we 
reported\footnote{\url{https://github.com/ConsenSys/eth2.0-dafny/issues}}, 
and also on the practicality of this kind of project.

\subsection{Array-out-of-bounds Runtime Error}\label{sec-array-out-of-bounds}

The function \verb+get_attesting_indices+ (Listing~\ref{algo-get_att}) is called from within several important components of the 
\texttt{state\_transition} function including the processing of rewards and penalties, justification and 
finalisation, as well as the processing of attestations (votes). 
%

The last line (13) of \verb+get_attesting_indices+ collects the indices in the array \verb+committee+
that have a corresponding bit set to true in array \verb+bits+ and returns it as a \emph{set} of indices.
%
The  length of \verb+bits+, noted $|\mathtt{bits}|$, is \verb+MAX1+.
Consequently, the following relation must be satisfied to avoid an array-out-of-bounds error:
$|\mathtt{committee}| \leq \mathtt{MAX1}$.
It follows that to prove\footnote{In Dafny, this check is built-in so you cannot avoid this proof.} the absence of array-out-of-bounds error in Dafny, the specification of \verb+get_attesting_indices+ (in Dafny) requires a pre-condition,  $|\mathtt{get\_beacon\_committee(\dots)}| \leq \mathtt{MAX1}$ (line~10).

\begin{lstlisting}[language=python3,xleftmargin=2em,label=algo-get_att,caption={Python code for \texttt{get\_attesting\_indices}.}]
    def get_attesting_indices(
        state: BeaconState, 
        data: AttestationData, 
        bits: Bitlist[MAX1]
    ) -> Set[ValidatorIndex]:
    """
    Return the set of attesting indices corresponding to 
    ``data`` and ``bits``.
    """
    committee=get_beacon_committee(state, data.slot, data.index)
    return 
        # Collect indices in committee for which bits is set 
        set(index for i, index in enumerate(committee) if bits[i])
    \end{lstlisting}
    
This pre-condition naturally imposes a post-condition for \verb+get_beacon_committee+ and
trying to prove this post-condition we uncovered a very subtle bug: depending on the number of \emph{active validators} $V$ in \verb+state+:
\begin{description}
    \item[$\mathbf{V \leq 4,194,304}$:] there is no array-out-of-bounds error as 
    $|\mathtt{get\_beacon\_committee(\dots)}| \leq \mathtt{MAX1}$ 
    for \emph{all} input combinations of \verb+data.slot+ and \verb+data.index+,
    \item[$\mathbf{4,194,304 < V < 4,196,352}$:] there is at least one input combination of \verb+data.slot+ and \verb+data.index+
    that triggers an array-out-of-bounds with $|\mathtt{get\_beacon\_committee(\dots)}| > \mathtt{MAX1}$, and
    \item[$\mathbf{4,196,352 \leq V}$:] for all input combination of \verb+data.slot+ and \verb+data.index+, 
    there is an array-out-of-bounds  
    $|\mathtt{get\_beacon\_committee(\dots)}| > \mathtt{MAX1}$. 
\end{description}

This previously undocumented bug was difficult to detect. It required many hours of effort to model the dynamics of the
problem; the analysis was quite complex due to the multiple interelated parameter calculations, and the use
of floored integer division. The full description and the analysis of this bug has been reported as issue\footnote{\url{https://github.com/ethereum/consensus-specs/issues/2500}} to the 
reference implementation github repository. 
The issue was confirmed by the reference implementation writers.

\subsection{Beyond Runtime Errors}
We have also been able to establish some well-formedness properties of the data structure that represents the \emph{block-tree} built by each node.
Each added block has a stamp, the \emph{slot number} and a link to its \emph{parent}.
The block-tree is the tree representation of the parent relation.
The block-tree should satisfy the following properties
\footnote{In the \texttt{ForkChoice.dfy} file as \href{https://github.com/ConsenSys/eth2.0-dafny/blob/4e41de2866c8d017ccf4aaf2154471ffa722b308/src/dafny/beacon/forkchoice/ForkChoice.dfy\#L203}{list of invariants} on the \texttt{Store}.}:
\begin{itemize}
    \item Every block $b$ except the genesis block has a parent,
    \item Every block $b$ with parent $p$ is such that the slot of $b$ is strictly larger than the slot of $p$,
    \item the transitive closure of the parent relation produces chains of blocks that are totally ordered using the $<$ relation on slot,
    \item  the smallest element of each chain has slot 0 (and consequently is the genesis block).
\end{itemize}

Another noticeable contribution compared to other approaches (like testing) is that we have proved the termination of all loops.

\subsection{Finalisation and Justification}
During the course of the project we benefited from the guidance of the third co-author who has comprehensive expertise in various aspects of the Beacon Chain, including the \emph{fork choice} part, and identified the  \emph{fork choice} implementation of the reference implementation as a component that needed verification. 

The  \emph{fork choice rules} are designed to identify a \emph{canonical branch} in the block-tree which in turn defines the \emph{canonical chain}.
To achieve this goal, we first assumed a fixed set of validators.
Then we built a Dafny proof of the GasperFFG~\cite{gasperFFG} protocol and tried to prove properties about the \emph{justified and finalised blocks} in the block-tree.
We could mechanically prove Lemmas~4.11 and 5.1, Theorem~5.2 from~\cite{gasperFFG}. Note that a complete proof in Coq is available in~\cite{gasper-coq} but it does not use the Beacon Chain data structures.
%
We only managed to push these properties up to a certain level on the functional specifications of our code base and not on the actual reference implementation. Doing so would require to add a substantial amount of details and to modify the structure of several proofs which was not doable in our timeframe.
This experimental work is archived in branch \verb+goal1+ of the repository. 
There is ongoing work focussing on this topic: designing the mechanised proofs\footnote{\url{https://github.com/runtimeverification/beacon-chain-verification}} of the refinement soundness of the state transition function (Phase 0) w.r.t. the GasperFFG protocol.

\subsection{Reflection}

\paragraph{\it \bfseries Verification Effort.}
The net effort for formal verification took 16 person-months. This figure is for the 
\emph{Beacon Chain State Transition} and does not include the time spent on the SSZ and Merkleise 
libraries that were completed before this project started. The division of time was primarily between 
the second and third components of the project. Translation of the reference implementation in Dafny, 
took approximately 6 person-months\footnote{This translation includes the proof of absence of runtime errors.}. Synthesis of functional specifications, including proofs, took approximately 
10 person-months. The time allocation for the identification of simplifications is more difficult to assess.
Though some consideration was given initially, this aspect was ongoing, as our understanding of the reference 
implementation evolved. 

\paragraph{\it \bfseries Trust Base.}
The validity of the verification results assumes the correctness of the Dafny specification and the Z3 verifier.
Dafny is actively maintained and under continuous improvement. And in the rare instance where Dafny behaves
unpredictability, bug reports are responded to in a timely manner. During the course of this project a few bugs 
were reported. For example it was found that the definition of an inconsistent \texttt{const} could lead to unsound 
verification results and reported as an issue\footnote{\url{https://github.com/dafny-lang/dafny/issues/922}} (fixed) to the Dafny language github repository

\paragraph{\it \bfseries Practicality of the Approach.}
The use of Dafny does not require any specific knowledge beyond standard program verification (Hoare style proofs) and first-order logics. 
There is ample support (videos, tutorials, books) to help learning how to write Dafny programs and proofs.
The main difficulties/challenges in writing and verifying projects of this size with Dafny (and the same holds for other verification-friendly automated deductive verifiers) are: \textbf{1.} when the verification fails, it requires some experience 
to interpret the verifier feedback and make some progress,  
and \textbf{2.}  the unpredictability (time-wise) of the reasoning engine; this is due to the fact that \emph{verification conditions} that are generated by Dafny are in semi-decidable theories of the underlying  SMT-solver (Z3).  
In our experience, adding a seemingly innocuous line of proof may result in either a surge  or a drastic reduction   
of verification time.

\section{Conclusion}\label{sec-conclusion}

Overall this project was a significant undertaking. The complexity of the state transition mechanism, combined with 
the ambitous project scope, makes this one of the largest formal verification projects to be completed using
Dafny. Even with the model simplifications, that the Python language is not particularly compatible with the 
fundamentals that underpin formal verification presented continual challenges. Upon reflection: $i)$ the project would have benefited from a larger team and $ii)$ consideration of the application of formal verification methods 
earlier, ideally within the design process, would have had a positive impact.

The interest generated from this project provided an opportunity to facilitate Dafny training for the reference implementation writers
at the Ethereum Foundation. This training included the translation of code into Dafny, as well as more the
advanced topic of proof construction. Participants were able to gain insight into the formal verification process
which could provide valuable context when drafting future reference implementations and specifications.


\paragraph{\it \bfseries Acknowledgements.}

We thank the reference implementation writers at the Ethereum Foundation for their insightful feedback.
This project was supported by an Ethereum Foundation Grant, FY20-285 and we thank Danny Ryan (Ethereum Foundation) and Ben Edgington (ConsenSys) for their help in setting up this project and their support.
We also thank Roberto Saltini (ConsenSys) for his contribution at the early stage of the project.


\bibliography{beacon-bib}

\end{document}